# HOLISTIC RISK ASSESSMENT OF INFERENCE ATTACKS IN MACHINE LEARNING

Yang Yang[1]

[1] Donghua University, China
`2211824@mail.dhu.edu.cn`

**Abstract.**

As machine learning's expanding application, there are more and more unignorable privacy and safety issues. Especially inference attacks against Machine Learning models allow adversaries to infer sensitive information about the target model, such as training data, model parameters, etc. Inference attacks can lead to serious consequences, including violating individuals' privacy, compromising the intellectual property of the owner of the machine learning model. As far as concerned, researchers have studied and analyzed in depth several types of inference attacks, albeit in isolation, but there's still a lack of a holistic rick assessment of inference attacks against machine learning models, such as their application in different scenarios, the common factors affecting the performance of these attacks and the relationship among the attacks. As a result, this paper performs a holistic risk assessment of different inference attacks against Machine Learning models. This paper focuses on three kinds of representative attacks: membership inference attack, attribute inference attack and model stealing attack. And a threat model taxonomy is established. A total of 12 target models using three model architectures, including AlexNet, ResNet18 and Simple CNN, are trained on four datasets, namely CelebA, UTKFace, STL10 and FMNIST. Membership inference attack, attribute attack and model stealing attack are executed against these target models under different threat models respectively, and the performance of inference attacks is evaluated with attack accuracy as an indicator, which demonstrates that the complexity of the training dataset and overfitting level of the target models have a significant impact on the performance of inference attacks, while the effectiveness of the model stealing attack is negatively correlated with the membership inference attack.

**Keywords:** Machine Learning, Inference Attacks, Private Leakage, Risk Assessment.

## 1 Introduction

At present, machine learning is the basis of many popular internet intelligent services, such as image, voice recognition, recommendation system, etc. In the past few decades, significant progress has been made in machine learning, especially in the field of deep learning. The purpose of machine learning is to extract information from data through a prediction model, which is a function that can map feature vectors to categories or outputs. Among them, the deep learning model has become more and more popular because of its excellent generalization ability in many machine learning tasks, especially those involving computer vision and image recognition [1, 2]. The deep learning model is composed of many nonlinear mapping layers, from the input layer to the intermediate hidden layer and then to the output layer. Each connection between layers has a floating-point weight matrix as a parameter, and these weights are constantly updated in the training process. The topology between layers is related to the task of the model and plays an important role in the accuracy of the model.

However, with the more and more extensive use of machine learning models, more and more applications and services use deep learning algorithms to process a large number of user data. While providing great convenience to users distributed around the world, there are also serious problems of data disclosure or invasion of user privacy. For example, the patient's medical data is recorded and stored in the hospital's cloud service system for the hospital to analyze his health status, such as the probability that he may be infected with a certain disease. However, if these medical record data are leaked, it will cause great harm to the patient's privacy [3]. Because the training data of machine learning models are very sensitive, and the business value of the models themselves or their use in relevant security applications, these models are considered confidential. Attacks against machine learning models will bring serious security and privacy risks. Therefore, the research on the attack of machine learning model has attracted extensive attention of scholars at home and abroad.

In particular, deep neural networks are vulnerable to various inference attacks because their complex layer structure will remember the training data information. This paper focuses on the performance evaluation of inference attacks against machine learning. Inference attack is an attack that allows an adversary to speculate user data information from a target machine learning model. For example, information about training data, model parameters, etc. Inference attacks can lead to serious consequences, including threats to users' privacy, because machine learning models are often trained on sensitive data. In addition, inference attacks also may damage the intellectual property rights of the model owner.

This paper focuses on the classification task of supervised learning in machine learning. The training data of supervised learning is a set of correctly labeled data samples. And three representative speculative attacks are discussed: member speculative attack [4], attribute speculative attack [5] and model stealing attack [6]. The target of the first two attacks is the training data set of the model. The purpose is to determine whether a specific data sample belongs to the training data set, and recover it or a part of it, or predict the attributes unrelated to the initial task of the model. The model stealing attack involves the reconstruction of the non-public parameters of the target model.

In a word, at present, scholars have conducted in-depth research and Analysis on several speculative attacks, but they lack an overall assessment of the risk of speculative attacks, that is, they need to have a more comprehensive understanding of the risks caused by these attacks, such as the application scenarios of different attacks, the common factors affecting attacks, and the relationship between different attacks. In this paper, for the first time, the overall risk assessment of the security and privacy of the machine learning model is carried out for three representative speculative attacks.

This paper is divided into the following three parts:

(1) Threat model classification

This paper first classifies the background knowledge of the adversary according to two dimensions. One is the way to access the target model (black box / white box), and the other is the availability of auxiliary data sets (partial data sets, shadow data sets or no data sets). Focus on the three most popular speculative attacks and describe the different threat models they can apply. This provides a broader range of speculative attacks against machine learning models.

(2) Experimental evaluation

In this paper, machine learning models of three different model architectures (Alexnet [20], Resnet18 [21], SimpleCNN) are trained on four image data sets (CelebA [16], Fashion-MNIST [17], STL10 [18], UTKFace [19]), a total of twelve target models are trained. Three speculative attacks under different threat models are executed for these target models, and the performance of speculative attacks is comprehensively evaluated. The evaluation mainly analyzes the following four aspects: 1) the performance of different speculative attacks on different threat models and target models; 2) The influence of data set complexity on different speculative attacks; 3) The influence of model overfitting level on different speculative attacks; 4) Relationship between different speculative attacks.

3) Analysis of inference attack results

A large number of experimental results show that the complexity of the target model training data set has a significant impact on the accuracy of member speculative attack and model stealing attack. In particular, member speculation attacks on data sets with high complexity are more effective, but the model stealing attack is the opposite.

In addition, this paper also points out that there is a negative correlation between member speculative attacks and model stealing attacks, that is, when the target model is more vulnerable to member speculative attacks, it is less vulnerable to model stealing attacks. Compared with the shadow data set, the adversary can access part of the training data set, and the performance of member speculation attack, attribute speculation attack and model stealing attack is not significantly improved.

The following chapters are arranged as follows: Section 2 will introduce the background of machine learning and privacy risk; In Section 3, we introduce three algorithms of speculative attack; Section 4 evaluates the performance of speculative attacks in machine learning. Firstly, it introduces the model architecture and data set used by the target model and the relevant parameters of speculative attacks, and performs speculative attacks on the target model under different threat models; The results are analyzed with attack accuracy as an index, including evaluating the performance of attacks under different threat models, analyzing the impact of data set complexity and model overfitting level on speculative attacks, and the relationship between different attacks; At last, Section 5 summarizes the whole research.

## 2      Background and Threat Model Classification

This chapter will briefly introduce the relevant background of machine learning and privacy risk, and propose a threat model classification method, which classifies the threat models of speculative attacks according to two dimensions, i.e. 1) the access mode of the target model and 2) the auxiliary data set. Four different attack scenarios are considered.

### 2.1     Research Background

#### Machine Learning

*Learning Process.* Machine learning provides automatic analysis methods for large data sets to perform tasks that are difficult to be programmed. The general method of establishing machine learning model is divided into the following three steps [22]:

(1) Training: most machine learning models are regarded as a parameter function $h_\theta(x)$, the input is $x$, and the parameter is $\theta$. The input $x$ is usually represented as a set of vectors called features. The learning algorithm analyzes the training data to find out the value of the parameter $\theta$. In other words, in the training process, the purpose of machine learning algorithm is to learn the data characteristics related to the task. For example, given enough face images, the algorithm should learn to distinguish male and female images if it wants to recognize the gender of the person.

(2) Verification: model performance can be verified through a test data set. The test data set must not intersect with the training data set in order to test the generalization ability of the model. In other words, validation is actually a model integrity check to ensure that the algorithm has learned what it should learn.

(3) Prediction: once the training is completed, the model can be deployed to predict some inputs that are not in the training process. The parameter $\theta$ is fixed, and the model calculates $h_\theta(x)$ for the new input $x$. Model predictions may take different forms. For example, for classification problems, the most common is a set of probability vectors (confidence levels) that assign a probability to each category. Each element in the vector represents the possibility that the input belongs to this category.

*Optimization Algorithm.* The loss function $L(h, \theta)$ quantifies the cost of the difference between the predicted value $h_\theta(x)$ and the actual value $f(x)$ of the model, which is equivalent to the evaluation index of a model. Generally speaking, $h$ and $L$ are differentiable functions of $\theta$. The process of training the model $h$ is actually to find a value of $\theta$ on all training samples that makes the loss $L$ minimum or relatively small. In short, in machine learning algorithms, stochasticgradient descent and Adam are common optimization methods to maximize the accuracy of the algorithm in the process of machine learning model training. The models involved in this paper also use these two optimization algorithms, which will be described in detail below.

(1) Stochastic gradient descent (SGD): in this process, a single batch of data (possibly one or more samples) is continuously selected, the $h_\theta(x)$ and the corresponding $L(h, \theta)$ are calculated, and the $\theta$ value is adjusted in the opposite direction of the gradient to reduce $L(h, \theta)$, that is, the parameters are updated once, and the parameters are updated as many times as there are batches of data in the training round. Random means that the data selected for each update parameter is random, and the order of each round of sample update is also random. The advantage of this optimization method is that the parameter update speed is fast, but the disadvantage is that the amount of data used in each parameter update is small, resulting in large oscillation and instability during gradient update, and there may be large fluctuations in the optimal solution attachment.

(2) Adam: Adam algorithm is actually the combination of momentum algorithm and RMSprop [23]. Momentum refers to the momentum gradient descent algorithm. The concept of momentum is introduced into the original gradient descent algorithm. The direction of parameter update will be affected by the direction of the previous gradient, that is, each gradient update will have the relationship between the directions of the previous gradients. This algorithm makes the gradient change smoother and alleviates the oscillation problem to a certain extent. RMSprop is an improved adaptive learning rate gradient descent algorithm. During the training process, the learning rate can be automatically adjusted. Different parameter frequencies correspond to different adaptive learning rates, which enables the algorithm to update to different degrees according to the importance of different parameters. Adam algorithm absorbs the advantages of adaptive learning rate gradient descent algorithm and momentum algorithm, and improves the gradient calculation method and learning rate, making it an optimization algorithm with good performance.

*Task Classification.* According to the existing data structure, it can be divided into two main types:

(1) Supervised learning: training samples are input forms with corresponding output labels. If the output result is a category, it is a classification task. In this learning task, the purpose of the model is to map the input data samples to a category. Common implementation methods include support vector machine, logistic regression and neural network; If the output is an actual value, it is defined as a regression problem. Typical examples of supervised learning tasks include spam filtering and target recognition in images. Under this setting, the model parameters are continuously adjusted to reduce the gap between the model prediction $h_\theta(x)$ and the expected output $f(x)$ of the data set.

2) Unsupervised learning: if the given input is unmarked, this kind of task is the unsupervised learning task. Unsupervised learning considers clustering and reducing the dimension of project data to low dimensional subspace. Common algorithms include K-means, AP clustering and principal component analysis.

Other types, such as reinforcement learning, will not be discussed here.

*Model Training Mode.* According to whether the data are centralized or decentralized before model training, the machine learning model training mode can be divided into the following three categories [24].

(1) Centralized learning: in traditional machine learning, the training data sets used by each user are stored in the central server. The target model is trained in this joint pool. The advantages of this model training method are that it is easier to train and deploy the model, and the accuracy of the model is higher, so it is widely used in the actual scene; The disadvantage is that it increases the burden of computing resources and storage space of the central server, and exposes all user data to security and privacy risks. In other words, once user data is uploaded to the central server, it is difficult for users to understand and control it, and they do not know whether the data will be used or transmitted to unknown third parties.

(2) Distributed learning: the training data set and calculation load used by the user are distributed on each node. The central server only maintains global parameters to reduce the calculation burden. All nodes jointly train a machine learning target model. The central server occupies a dominant position, and the connection between each node and the central server is stable.

(3) Federated learning: Federated learning can be regarded as a special form of distributed machine learning. In this learning mode, each user has his own training data set, trains a local model on his own data set, but regularly exchanges and updates model parameters, or builds a partial model, and trains a joint model through the central server. The data is scattered during the training process. Federal average algorithm is one of the most popular methods in federal learning. In federated learning, users can independently control their own devices and data, and can also decide when to participate in or quit federated learning. For example, Google has deployed federated learning on millions of devices to train the keyboard to predict character sequences through user input on the mobile phone keyboard.

**Privacy Risk.** Cristofaro et al. [22] discussed the privacy problem in machine learning. In the past, suppliers such as Google, Microsoft and Amazon have started to provide software interfaces for customers, so that they can easily embed machine learning tasks into their applications. That is, these organizations can use mlaas (machine learning as a service) to outsource complex tasks, such as training classifiers, performance prediction, clustering, etc. Others can also query models trained based on their data. This method is also often used in some government cooperation or scientific projects. However, if some malicious users can recover the data of training these models, the information leakage will pose a serious threat. Similarly, if the internal parameters of the model are confidential, the adversary should not be allowed to obtain these parameters when accessing the model.

According to the different content of machine learning privacy protection, machine learning privacy can be divided into the following three categories [24]:

(1) training data privacy: refers to user data in machine learning, including personal identity information and sensitive data information. [20] The training data privacy is defined as that the adversary can learn or speculate information about the training data from the target model, but cannot obtain such information from other models trained with the same distribution of data sets.

(2) Model privacy: refers to the privacy information related to the machine learning model such as model training algorithm, model weight parameters, model topology, activation function and super parameters in machine learning.

(3) Prediction result privacy: prediction result privacy refers to the prediction result made by the model to the user's input request and the sensitive information that the user is unwilling to disclose in machine learning.

### 2.2 Threat Model Classification

**Access Target Model.** We consider two access methods: white box and black box. Under the setting of the white box, the enemy has all the information of the target model, including its parameters and structure. Under the access mode of black box, the adversary can only access the target model in the form of API. For example, they can query the target model and obtain the output of the model. However, most of the existing black box attacks [4, 12, 25] also assume that the adversary knows the structure of the target model and can be used to construct the shadow model.

In general, the white box attack model reflects those scenarios where the parameters of the target model are leaked. For example, under federated learning, multiple participants can share the parameter updates of the model and train a target model cooperatively. The black box attack model describes the encapsulated API access, such as the platform that provides machine learning as a service (mlaas). Users can only observe the prediction of the model.

**Auxiliary Data Set.** The adversary needs an auxiliary data set to train their attack model. In this paper, two kinds of auxiliary data sets are considered: 1) partial training data set and 2) shadow data set. In the first scenario, the adversary can obtain a part of the actual training data set of the target model, while in the second scenario, the adversary can obtain the data set with the same distribution as the target model training data set, that is, the shadow data set.

**The Scenario Sets.** Two types of model access methods and the availability of two auxiliary data sets to form four threat models, namely, black box / partial data set, black box / shadow data set, white box / partial data set and white box / shadow data set. Different speculative attacks can be applied in different scenarios.

## 3 Description of Speculative Attack Algorithm for Machine Learning Model

This chapter will introduce the three speculative attacks discussed in this article, namely, member speculative attack, attribute speculative attack and model stealing attack. The first two kinds of training data are designed to speculate the machine learning model, which pose a threat to the privacy of training data; The last purpose is to steal the parameters of the target model, that is, to infringe the privacy of the model. Each attack and each threat model focuses on a representative, state-of-the-art approach. In addition, this chapter will also introduce the machine learning target model attacked, including the data set and model architecture used for training.

### 3.1 Member Speculation Attack

The meaning of the speculative attack against the members of the machine learning model is [4]: a data sample is given, and the adversary judges whether it belongs to the training data set of the model. The purpose of the attacker is to establish an attack model that can identify the differences in the behavior of the target model, and distinguish the members and non members of the target model training data set based on the output of the target model [4]. The main method is to train an attack model to distinguish the behavior difference between the training input behavior and the untrained input behavior of the target model. The output of the target model is a probability vector, and each element represents the probability that the data belongs to the category. The output prediction vector and the class label of the data train the attack model together. In other words, the member conjecture attack is actually a classification problem, and the accuracy of its classification directly indicates the degree of leakage of the model to its training data. The schematic diagram is shown in Fig. 1. The following describes how to implement member speculation attacks under different threat models.

Speculating whether the target sample is a member of the model training data set will cause a serious privacy threat; For example, if a machine learning model used to predict which drugs a patient takes uses a patient with a certain disease as a training data set, the member speculates that the attack will reveal the health status of the individual. Members speculate that the appearance of an attack is often a signal indicating that the target model will lead to potential privacy leakage, and is also a channel for further attacks [7].

Members speculate that the attack can be launched against the target model in two attack access modes of black box and white box. Under the black box setting, using the method proposed by Salem et al. [7], the attack model has two inputs: the posterior value of the target sample ranking and the binary index of whether a target sample is correctly predicted. Each input is first input to a different two-layer MLP (multilayer perceptron), and then the resulting two embedded connections are input to a four layer MLP.

(1) Black box / shadow data set this is the most common and difficult attack scenario [4,7]. The adversary has black box access to the target model and has a shadow auxiliary data set with the same distribution as the actual training data set of the target model. The enemy's attack can be divided into three steps [7]. The first is to train the shadow model. The adversary divides the shadow data set into two parts, and uses one of them to train a unique shadow model with the same task as the target model. The data sets of the target model and the shadow model are equally distributed but do not intersect. Then, the adversary uses the trained shadow model to predict all the data samples (including the shadow training data set and the shadow test data set) in the entire shadow data set. For each query sample, the shadow model returns its posterior value and prediction category. If the sample is part of the shadow model training data set, the adversary marks it as a member, otherwise it is a non member. Then the adversary trains an attack model, that is, a binary classifier, with the data set composed of the obtained posterior values and the corresponding markers. Finally, when the sample is input to the target model, the output prediction vector and prediction category are input to the attack model, so as to judge whether the data sample belongs to the training data set of the target model.

(2) Black box / partial data set: the adversary can access the target model in black box and has part of the training data set. Its attack method is similar to that of black box / shadow data set [7]. However, the adversary no longer needs to train the shadow model, but directly trains the attack model based on some training data sets.

Under the white box setting, referring to the method proposed by Nasr et al. [26], the attack model has four inputs, including the posterior value of the target sample ranking, the classification loss, the parameter gradient value of the last layer of the target model, and the unique hot coding of its real tag. Each input is sent to a different neural network, and the embedding of its output is connected together as the input of the four layer MLP.

(1) White box / shadow data set Nasr et al. [26] introduced an attack with shadow or partial training data set under white box setting. The former is similar to the black box / shadow data set. The adversary uses the shadow data set to train a shadow model that mimics the behavior of the target model and generates data to train the attack model. Because the adversary can access the target model in a white box, they can also use the gradient value of the target sample relative to the model parameters, the embedding from different middle layers, the classification loss and the predicted posterior value and category.

(2) The attack method under the threat model of white box / partial data set is almost consistent with the corresponding black box method [26]. The only difference is that the adversary can use the same feature set as the white box / shadow data set attack model.

### 3.2 Attribute Speculation Attack

The machine learning model may learn additional information about the training data unrelated to its initial task, that is, the adversary can infer the attribute of the training data, which is independent of the attribute that the target model wants to capture. For example, a face recognition model can also learn to predict its race. In modern deep learning models, all features will be individually represented, and some attribute inference attacks unrelated to the main task of the model are intended to use this inadvertent information disclosure to infer other non target attributes unrelated to the task of the target model.

Currently, the most advanced attack methods mainly rely on the embedding of target samples from target models to predict the target attributes of samples [5,13], which requires the adversary to have white box access to the target models. More specifically, in the training process, participants provide a batch of training data for each iteration, and the adversary can infer the attribute of a single batch from it, that is, the adversary can infer that the data in a given batch has the attribute, but not in other batches. In addition, the adversary can also infer when the content appears in the training data, which has a very serious impact on privacy.

Attribute speculation attacks can be applied under the two threat models of white box / partial data set and white box / shadow data set, and their attacks follow similar attack methods [5,13]. The only difference is the data set used to train the attack model. In both cases, it is assumed that the adversary knows the target attribute of the auxiliary data set. Then they use target model embedding and target attributes to train a classifier to launch an attack.

### 3.3    Model Stealing Attack

Model stealing attack, also known as model extraction attack, aims to extract the parameters of the target model [6,14]. In this attack, the adversary has black box access rights, but does not have any knowledge about machine learning model parameters or training data. The schematic diagram of machine learning model extraction attack is shown in Figure 3. In an ideal situation, the adversary can obtain a theft model with very similar performance to the target model, such as nearly 100% consistency in input. The model stealing attack actually takes advantage of the richness of the output information of the machine learning prediction API, such as high-precision confidence (posterior value) and class labels.

Model theft attacks can cause serious privacy risks. For example, it is often difficult to train a high-level machine learning model due to the lack of data or computing resources. Stealing a trained model essentially constitutes intellectual property theft. And in many other attacks, such as adversarial samples [27], the target machine learning model is required to have white box access rights. At this time, the model stealing attack is actually equivalent to providing a path for the execution of these attacks.

Model theft attack can be applied under two threat models, black box / partial data set and black box / shadow data set. This paper uses the attack method proposed by tramer et al. [6]. Assuming that the adversary knows the structure of the target model, the two attack scenarios follow similar attack methods. The adversary uses the data samples of its auxiliary data set (part / shadow) to query the target model to obtain the corresponding posterior value. They then use these posterior values as a factual basis to train the theft model.

### 3.4    Machine Learning Target Model

This paper focuses on one of the most popular machine learning applications, image classification. Usually, the goal of machine learning classifier is to map a data sample to a category. The input is the data sample and the output is the confidence. It is a probability vector. Each element in the vector represents the possibility of belonging to a category.

Training machine learning target model as the target of speculative attack, this paper selects 4 data sets and 3 advanced machine learning models to train a total of 12 target models, and then launches three speculative attacks on the target models, and returns the attack accuracy, so as to comprehensively analyze and evaluate the results of different speculative attacks. The data set and model architecture used in the target model training will be described below.

**Datasets.** The following four data sets were used for the experiment.

CelebA includes 202599 face images, and each image is related to 40 binary attributes [16]. We select and combine 3 attributes, including heavy makeup, slightly swollen mouth and smile, to form 8 categories of the target model.

Fashion MNIST (fmnist) is a data set with about 70000 gray-scale images, whose images are evenly distributed in 10 different categories, including T-shirts, pants, pullovers, dresses, coats, shirts, sports shoes, handbags, sandals and boots [17]. Fmnist dataset has the lowest complexity because it only contains gray-scale images.

STL10 is an image data set with 10 categories, and each category contains about 1300 images [18]. The category consists of aircraft, birds, cars, cats, dogs, deer, horses, monkeys, ships and trucks. The STL10 dataset is the most complex of the four datasets because it contains 10 different categories.

UTKFace has 23000 images associated with age, gender and race [19]. This paper considers the four largest races in the data set (black, white, Asian and Indian) and uses race as the category of the target model. A total of 22012 images for the four races.

The sample size in all data sets was adjusted to 32 * 32 pixels. This is common in machine learning, ensuring that comparisons between different data sets are fair. Each data set is randomly divided into the following four equal and disjoint parts, including target training data set, target test data set, shadow training data set and shadow test data set.

(1) Target training data set: this data set is used to train all target models and evaluate the performance of member speculative attacks. For speculative attacks under the threat model that requires part of the training data set, 70% of the samples of this data set are randomly selected as part of the training data set.

(2) Target test data set: this data set is used to evaluate the performance of the target model. It is also used to evaluate the performance of member speculation, attribute speculation and model theft attacks.

(3) Shadow training data set: this data set is used to train all attack models that require a shadow auxiliary data set.

(4) Shadow test data set: this data set is used to train two member speculative attacks that require shadow data set threat model, namely black box / shadow and white box / shadow.

In summary, all data sets selected in this paper are benchmark data sets used to evaluate speculative attacks against machine learning models. These data sets have different categories and cover a wide variety of objects, such as faces, vehicles and animals. In addition, the image includes both a gray-scale image and a color image.

**Target Model Architecture.** The target model uses the following three popular model architectures:

AlexNet is a convolutional neural network with 60 million parameters and 650000 neurons, consisting of five convolution layers, some of which are followed by the maximum pooling layer, three full connection layers, and the last layer is the softmax layer of 1000 channels [20]. The activation function is ReLU function, and the regularization method of dropout is introduced to solve the overfitting problem that is easy to occur in the model training process, so that the performance of the deep convolu-tion neural network is significantly improved. This is a classical neural network model, which shows superior performance in image classification. At the same time, due to the use of dropout, it is relatively difficult to have over fitting problems.

Resnet18 uses residual blocks to build a residual network learning framework [21]. The residual block redefines the layer as a fitting residual function according to the layer input, i.e. The input can be propagated forward faster through the residual connection between layers. This residual mapping is often easier to optimize and makes the deep neural network easier to train. There are 18 weight layers, including convolution layer and full connection layer. Compared with the other two neural networks, resnet18 is characterized by its deep network, long model training time and high model accuracy.

SimpleCNN is a simple convolutional neural network built manually, consisting of two convolution layers and two full connection layers.

The three machine learning model structures are used for training on the four data sets mentioned above. A total of 12 different target models were trained. This is the simplest of the three neural networks. The number of network layers is small and the model training is fast.

When training the target model, the batch size is set to 64, and the cross entropy is used as the loss function[27,28]. We used SGD as the optimization method, with a weight delay of 5e-4 and a momentum of 0.9. Each target model was trained for 100 rounds. The learning rate was set to 1e-3, and table 2 lists the training / test accuracy of all target models. Note that the training method of the shadow model required by the member's speculative attack is the same as that of the target model, except that the training data set uses the shadow test data set[29, 30, 31, 32].

## 4 Performance Evaluation of Speculative Attacks Against Machine Learning Models

In this chapter, we first evaluate the overall performance of four speculative attacks, with accuracy as the evaluation index; Then, the influence of data set complexity and overfitting on attack performance and the relationship between different attacks are analyzed. Specifically, this part of the work aims to answer the following four key questions:

Q1: what is the performance of speculative attacks under different threat models?

Q2: how does the complexity of the training data set affect the performance of three different speculative attacks?

Q3: what is the impact of target model overfitting on the performance of three different speculative attacks?

Q4: what is the relationship between the three different speculative attacks?

### 4.1 Performance Evaluation of Speculative Attacks Under Different Threat Models

Taking accuracy as an indicator, the histogram is drawn under four data sets, three model architectures and different applicable threat models. Observe and analyze the performance of speculative attacks.

Figure 4, figure 5 and Figure 6 respectively show the performance of member speculative attacks under different threat models with different data sets and target model architectures of alexnet, resnet18 and simplecn. It can be observed that this attack has high accuracy on celeba, stl10 and utkface datasets. When using the stl10 dataset, the performance of member speculation attack is the best. For example, in the target model with the structure of stl10 data set resnet18, the attack accuracy is as high as 0.929 under the attack scenario of white box / part. However, the performance of member conjecture attack is not ideal on fmnist data set, because the model trained on fmnist has strong generalization ability in non member data samples, that is to say, its degree of overfitting is low[33, 34, 35].

When resnet18 is used as the target model architecture, the performance of member speculative attacks is relatively the best. However, using the target models of alexnet and simplecnn, there is no significant difference in the performance of their members' speculative attacks. This is because resnet18 has a large number of neural network layers, and it is easier to over fit during training. Members speculate that attacks can obtain better attack performance by using this feature[36, 37, 38, 39].

In addition, the adversary of white box access to the target model can usually obtain better attack performance than black box access. For example, members speculate that the attack accuracy under the white box / shadow setting is higher than that under the black box / shadow setting, and the same phenomenon also appears in the white box / part and the black box / part. This is expected because the adversary can use more information under the white box setting. Moreover, the member speculation attack performance using part of the training data set is also better than that of the shadow data set, but it is not significant[40, 41, 42].

### 4.2 Influence of Dataset Complexity on Performance of Different Speculative Attacks

In order to answer the second research question, for three different speculative attacks, the relationship between the complexity of the data set and the attack performance is drawn, as shown in FIGS. 13, 14 and 15. The abscissa represents four different data sets, the ordinate represents the speculative attack performance with accuracy, and each node in the graph is a speculative attack against a target model. For each speculative attack, the figure shows only one threat model.

As described above, all data samples in the four data sets were adjusted to 32 * 32 pixels. Fmnist is the simplest dataset, because it only contains gray-scale images, then utkface containing full-color faces, then celeba, which contains 10 times more images than utkface, and the most complex is the stl10 dataset, which contains 10 different categories of images from cats to ships.

## 5    Conclusion

It is speculated that the attack will pose a serious threat to the user's privacy information. For the first time, this paper makes a comprehensive evaluation and analysis of the privacy risk caused by three speculative attacks against machine learning models under different threat models. In addition, a threat model classification method is established in terms of the way to access the target model and the data set for the three speculative attacks of member speculation attack, attribute speculation attack and model theft attack. The model architecture and four data sets as well as applicable threat models have been extensively studied. Through the analysis and observation of the experimental results, we find that the complexity and overfitting level of the training data set play an important role in the performance of the attack. The more complex the data set, the better the effect of the member's speculative attack, and the worse the effect of the model stealing attack; The overfitting of the target model will also promote the member speculation attack and suppress the model stealing attack. It has no obvious effect on attribute speculation attacks. At the same time, the attack performance of model stealing attack and member speculation attack is negatively correlated.

At present, there are several effective defense mechanisms against speculative attacks. For example, differential privacy and knowledge distillation still lack comprehensive quantitative research and evaluation on whether the defense mechanism can effectively protect the machine learning model and mitigate speculative attacks. The follow-up work can launch speculative attacks on the machine learning target model using the defense mechanism, observe and compare the performance of speculative attacks, and evaluate the effectiveness of the defense mechanism against speculative attacks.